\documentclass{svmult}
\usepackage[utf8]{inputenc}
\usepackage{amsmath}
\usepackage{algorithm}
\usepackage{mathptmx}
\usepackage[]{algpseudocode}
\usepackage{tikz}
\usepackage{tabularx}
\usepackage{caption}
\usepackage{enumitem}

\author{Ivan S. Grechikhin}

\title*{Comparison of statistical procedures for Gaussian graphical model selection}

\algnewcommand{\LineComment}[1]{\State \(\triangleright\) #1}
\algnewcommand\textfunc{\textsc}

\begin{document}


\maketitle\footnotetext{National Research University Higher School of Economics, Laboratory of Algorithms and Technologies for Network Analysis, 136 Rodionova St, Nizhny Novgorod 603093, Russia}

\abstract{Graphical models are used in a variety of problems to uncover hidden structures. There is a huge number of different identification procedures, constructed for different purposes. However, it is important to research different properties of such procedures and compare them in order to find out the best procedure or the best use case for some specific procedure. In this paper, some statistical identification procedures are compared using different measures, such as Type I and Type II errors, ROC AUC. \\}


\section{Introduction}

Graphical models are used as an applications in various areas of science such as bioinformatics, economics, cryptography and many more \cite{drton, jordan}. The main reason for extensive use of graphical models is the level of visualization, which allows a scientist to uncover hidden patterns and connections between entities. However, the reliability of the obtained network is under discussion. In this paper gaussian graphical models are discussed. 

Usually, gaussian graphical model is a graph, where every vertex is a random variable and the random vector is distributed according to the multivariate normal distribution\cite{drton}. Connections in the graph show some kind of dependency between two variables. There are three types of graphs widely used. First of them is bi-directed graph; in this model, two random variables are marginally independent if there are no edge between corresponding vertices. Such graphs are usually constructed on correlation matrices, where zero correlation means marginal independence or absence of the edge in graph. Other case of independence is conditional independence on all other variables: undirected graphs are used for these patterns. Undirected graphs are constructed from concentration or partial correlation matrices, which are just inverses of covariance matrix; zero element of the concentration matrix means lack of edge in the graph between two corresponding vertices. Finally, directed acyclic graphs display conditional independence on a subset of random variables. 

In the literature, there are some methods or approaches for statistical inference. Drton \& Perlman\cite{drton} suggested some methods of family-wise error rate control, using different p-value adjustment. Those methods control the pre-determined level of family-wise Type I error (the probability of at least one Type I error in the model). 

Also, Drton \& Perlman \cite{sinful} suggested SINful procedure as a way of obtaining more conservative and less conservative networks. Later, this procedure was improved by introducing Minimal Covariance Determinant estimator \cite{gottard}, because it makes the SINful procedure more robust. However, the authors do not suggest any algorithmic procedure to find the boundaries of significant, indeterminate and non-significant p-values.

Another approach tries to obtain the best final network or correlation matrix corresponding to the graphical model using some function, which estimates the score of a model\cite{schafer, bayescovlasso, lasso1}. Basically, this approach comes in part from optimisation field. Recently, L1-regularization for optimisation function gains popularity in probabilistic inference field\cite{bayescovlasso, lasso1}. The reason behind this is the fact that lasso regularization has a property, where great part of the parameters in optimisation function become zero. This fact helps to distuinguish zero correlation coefficients from non-zero values and obtain network, where edges connect correlated variables. This regularization particularly helps with sparse correlation matrices, where a lot of elements are assumed to be zero.

In the article, undirected gaussian graphical models will be discussed. Statistical methods of probabilistic inference are analysed. These methods are described by Drton \& Perlman \cite{drton}. The authors conducted experiments on procedures with different p-value adjustments. They showed that step-down adjustments on practice assimptotically achieve values of Family-Wise Error Rate close to the theoretical control level with increasing number of observations. On the contrary, other procedures on practice show decreasing level of FWER with increasing number of observations; their theoretically controlled FWER level is far from practically achieved. At the same time, Drton and Perlman do not analyze Type II errors. 

The goal of this article is to analyze properties of described in the Drton \& Perlman's article statistical procedures. Additionally to FWER, Type II errors, ROC AUC and risk functions will be compared. Type II error is useful to estimate total number of errors in a model or, in other words, the difference between true model and obtained model. ROC AUC allows to compare different models and analyze them without connection to the significance level. One more procedure will be added for comparison, which will be called simultaneous multiple testing procedure, which is optimal for risk function in the class of all unbiased procedures \cite{risk}.

\section{Undirected gaussian graphical models}

In this article, we concentrate on undirected gaussian graphical models. We have random vector $Y = (Y_1,..., Y_n)$, which is distributed according to the multivariate gaussian distribution $N_p(\mu, \Sigma)$ - hence the word "gaussian" in before "graphical models". Graph $G = (V,E)$ represents network that is constructed using the information from random variables.

The set $E$ of edges represents conditional independencies through Markov properties. It means that if the edge $(i,j)$ is absent from the set $E$, then two corresponding random variables from the random vector are conditionally independent, where the condition is induced on all other variables:
\begin{equation}
Y_i \parallel Y_j \hspace{4pt} | \hspace{6pt} Y_{V\\\{i,j\}}
\end{equation}
This pairwise Markov property or the conditional independence of two random variables also corresponds to the zero element of the concentration matrix. It is obtained from covariance matrix $\Sigma$ by inversion. This matrix can be coerced to the partial correlation matrix as well: if the elements of the concentration matrix are $\{\sigma^{ij}\}$, then:
\begin{equation}
\rho^{ij} = \frac{\sigma^{ij}}{\sqrt{\sigma^{ii} * \sigma^{jj}}}
\end{equation}
Here is an example of the graph constructed on some concentration matrix. 
\begin{center}
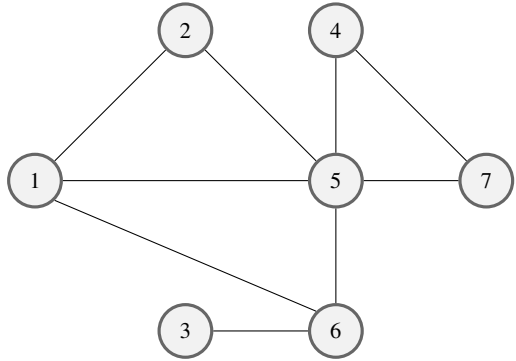
\begin{figure}
\centering
\begin{tikzpicture}[roundnode/.style={circle, draw=black!60, fill=black!5, very thick, minimum size=7mm}]

\node[roundnode] (first) at (0,2) {1};
\node[roundnode] (second) at (2,4) {2};
\node[roundnode] (third) at (2,0) {3};
\node[roundnode] (fourth) at (4,4) {4};
\node[roundnode] (fifth) at (4,2) {5};
\node[roundnode] (sixth) at (4,0) {6};
\node[roundnode] (seventh) at (6,2) {7};

\draw (first.east) -- (fifth.west);
\draw (first.north east) -- (second.south west);
\draw (first.south east) -- (sixth.north west);
\draw (second.south east) -- (fifth.north west);
\draw (third.east) -- (sixth.west);
\draw (fourth.south east) -- (seventh.north west);
\draw (fourth.south) -- (fifth.north);
\draw (fifth.south) -- (sixth.north);
\draw (fifth.east) -- (seventh.west);

\end{tikzpicture}
\captionsetup{justification=centering}
\caption{Example of a graph, constructed with concentration matrix}

\end{figure}
\end{center}
\begin{table}[h]
\centering
\begin{tabular}{| p{0.5cm} | p{0.8cm} | p{0.8cm} | p{0.8cm} | p{0.8cm} | p{0.8cm} | p{0.8cm} | p{0.8cm} | }

\hline
$\Sigma$ 	& 1 		& 2 		& 3 		& 4 		& 5 		& 6 		& 7 	\\ \hline
1 		& 1 		& 0.465 	& 0 		& 0 		& 0.511	& 0.392	& 0 	\\ \hline
2 		& 0.465 	& 1 		& 0 		& 0 		& 0.448	& 0 		& 0 	\\ \hline
3 		& 0 		& 0 		& 1 		& 0 		& 0 		& 0.32	& 0 	\\ \hline
4 		& 0 		& 0 		& 0 		& 1 		& 0.262	& 0		& 0.314 	\\ \hline
5 		& 0.511	& 0.448	& 0 		& 0.262	& 1 		& 0.459	& 0.42 	\\ \hline
6 		& 0.392	& 0 		& 0.32	& 0 		& 0.459	& 1 		& 0 	\\ \hline
7 		& 0 		& 0 		& 0 		& 0.314	& 0.42	& 0 		& 1 	\\ \hline
\end{tabular}
\caption{Example of concentration matrix}
\end{table}

If the dimensionality of the random vector is $p$, then the number of edges is $P = \frac{p*(p-1)}{2}$. For every edge, there is a hypothesis:
\begin{equation}
h_{ij}: Y_i \parallel Y_j \hspace{4pt} | \hspace{6pt} Y_{V\\\{i,j\}} \hspace{8pt} or \hspace{8pt} \rho^{ij} = 0
\end{equation}
against alternative
\begin{equation}
k_{ij}: Y_i \not\parallel Y_j \hspace{4pt} | \hspace{6pt} Y_{V\\\{i,j\}} \hspace{8pt} or \hspace{8pt} \rho^{ij} \ne 0
\end{equation}
As a result, there are $P$ different hypothesis to test to determine the structure of the graph. The hypothesis is rejected if the probability of it being true, based on the existing data is too small, less than some significance level $\alpha$, chosen beforehand.

There are a lot of procedures, which are described in the literature for constructing graphs on networks on data. Usually, the exact steps of procedures depend on the end goal, however, we can distinguish three common types of procedures. The first type might be called statistical because these procedures rely on some statistical properties of source data in order to find the network. In particular, different identification procedures of this type are suggested by Drton \& Perlman\cite{drton}. Some properties of these statistical identification procedures will be observed in the paper. 

Other types of procedures include procedures that optimise some goodness-of-fit function, that is based on the graph structure. Some of procedures use Bayesian approach, which is considered more demanding, because it needs to know prior information about distribution and compute posterior distribution. However, in this paper only properties of statistical procedures will be observed.

The goal of any procedure is to uncover the underlying patterns and connections of random variables. It means that there are basically two kinds of networks. True network or true correlation matrix defines how random variables are connected in reality, however we usually do not know the true network structure. Therefore we use data from observations on random variables, and from that data we construct sample correlation matrix or sample network. This network represents some graph, which is considered to be close to the true network, however the data may have provided us with some misleadings. As a result there are errors that we can consider as a measure of uncertainty in the network.  

First of all there are two types of errors: when we add edge, that do not exist in the true network and when we miss the edge that exist in true graph. The Type I Error is an error, when hypothesis is rejected, however it is true in reality. In our case Type I Error is when we consider establishing an edge between two edges, whereas in reality, the edge does not exist. Other way round, the Type II Error is when we accept a hypothesis, which is not true in real case. That means that we do not establish edge that exist in the true network. Based on these two types of errors, there are a lot of different measures of uncertainty.  

The most simple measure is a number of Type I or Type II errors, or in other words, the total number of wrongly added or wrongly missed edges. Sometimes errors of Type I are called True Positive (FP) and errors of Type II are called False Negative (FN), in that case number of correctly allocated edges is called True Positive (TP), and number of correctly absent edges is True Negative (TN). One of the most popular measures is Family-Wise Error Rate (FWER), which is a probability of making at least one Type I error in the whole network or $FWER = P(FP>0)$. False Discovery Rate (FDR) is a relation between the number of Type I errors and total number of edges in a sample network, or $FDR = \frac{FP}{FP + TP}$. Also known measures of errors include so called risk function, which is a linear combination of the number of Type I and Type II errors:
\begin{equation}
R(Y,\alpha) = E(FP)*(1-\alpha) + E(FN)*\alpha
\end{equation}

Additionaly, in this paper, we consider area under receiver operating characteristic (ROC) curve (ROC AUC). This curve is constructed for some procedure as follows: for any significance level, we calculate two characteristics: specifity, which is $Spe = \frac{TN}{FP+TN}$, and sensitivity, $Sen = \frac{TP}{TP+FN}$. We draw all points in two-dimensional space, where $X = 1-Spe$ and $Y = Sen$. As a result, when the significance level is zero, we will not reject any hypothesis, because there is no probability less than zero, $TP = 0$, $FP = 0$ and resulting point on a plot is $(0,0)$. On the contrary, if significance level is 1, using the same reasoning, all hypotheses will be rejected, $TN = 0$, $FN = 0$, as a result the point will be $(1,1)$. Resulting curve will be drawn from $(0,0)$ to $(1,1)$. If on some interval the curve is situated under the $y = x$ line, the output of the procedure might be inverted on the interval, which will only improve the procedure, however, optimal procedures are expected to be located on the upper side of $y = x$ line. Naturally, one can estimate the area under this whole curve, if it is equal to one, the procedure is optimal and ideal for any significance level; if area under curve is close to 0.5, the procedure practically does not differ from random decision. To sum up, ROC AUC allows to estimate the efficiency of the procedure as a whole, at different significance levels. 

\section{Identification procedures}

All described measures allow us to compare different procedures. The procedures are described below.

First of all, if we have observations $Y^{(1)},...,Y^{(n)}$ from a given multivariate normal distribution, where each observation is a vector of length $p$. Sample correlation matrix can be derived from observations using sample covariance matrix and sample mean:
\begin{equation}
S = \frac{1}{n-1}\sum_{m=1}^{n}(Y^{(m)}-\overline{Y})(Y^{(m)}-\overline{Y})^T
\end{equation}
\begin{equation}
\overline{Y}=\frac{1}{n}\sum_{m=1}^{n}Y^{(m)}
\end{equation}

Usually, for statistical procedures one needs to compute p-values. P-value is essentially a probability of trueness of hypothesis, given the current observations. The distribution of sample correlation coefficient, when true correlation coefficient is zero is known for components of normal random vector: if $r_{ij}$ is such sample correlation coefficient, then $ \sqrt{n-2}*r_{ij}\sqrt{1-r_{ij}^2}$ has a t-distribution with $n-2$ degrees of freedom. For sample correlation coefficient $n$ should be changed on $n-p$. Knowing this, we can obtain p-values for every sample correlation coefficient.

The simultaneous multiple testing procedure is the simplest one to obtain the network. In this procedure every hypothesis is tested independently at the same time with some chosen significance level. It means, that we compare the significance level with p-value for the hypothesis, if the p-value is lower, then the hypothesis is rejected and vice versa. It is worth noting that this procedure only controls the level of error in every hypothesis, but not in the whole network, which may be useful in some cases.

Drton \& Perlman\cite{drton} described procedures that control FWER in a network, which adjust p-value for every hypothesis. In this paper, four different adjustments are observed:

Bonferroni adjustment: 
\begin{equation}\pi_{ij}^{Bonf}=\min{\{C_p^2*\pi_{ij}, 1\}}\end{equation}

Sidak adjustment: 
\begin{equation}\pi_{ij}^{Sidak}=1 - (1-\pi_{ij})^{C_p^2}\end{equation}

If we reorder p-values in such a way that $\pi_{(1)}\leq\pi_{(2)}\leq...\leq\pi_{(C_p^2)}$, then

Bonferroni adjustment with Holm step-down procedure: \\
\begin{equation}\pi_{(a)}^{Bonf.Step} = \max_{b = 1,...,a}\min\{(C_p^2-b+1)*\pi_{(b)},1\}\end{equation}

Sidak adjustment with Holm step-down procedure: \\
\begin{equation}\pi_{(a)}^{Bonf.Step} = \max_{b = 1,...,a} 1 - (1-\pi_{(b)})^{C_p^2-b+1}\end{equation}

\section{Experiments and Results}

In the article by Drton \& Perlman\cite{drton}, they conducted some experiments to show that described procedures control FWER at pre-determined level. In the experiments they used generated concentration matrix with $p=7$, which has 9 non-zero elements from the interval $[0.2, 0.55]$. The number of observations for one trial varied from 25 to 500. Their experiments showed that step-procedures with Bonferroni and Sidak adjustments are closing in to pre-determined significance level $\alpha$. However, non-step procedures with the same adjustments are going further away from the pre-determined level, and the real FWER controlling level for those procedures is much lower than chosen $\alpha$.

First goal of the experiments was to repeat described experiments for matrices of higher dimensionality($p = 25$) and different concentration matrix densities($q=0.2,0.4,0.6,0.8,0.95$). Number of observations were chosen from 100 to 500 with step 100.
\begin{table}[h]
\centering
\begin{tabular}{| p{2cm} | p{0.8cm} | p{0.8cm} | p{0.8cm} | p{0.8cm} | p{0.8cm} | }

\hline
Bonferroni 		& 100 		& 200 		& 300 		& 400 		& 500 	\\ \hline
$E(P(FP>0))$ 		& 0.069 		& 0.057 		& 0.063 		& 0.058 		& 0.06	\\ \hline
$E(P(FN>0))$ 		& 1 			& 1 			& 1 			& 1 			& 1		\\ \hline
$E(FP>0)$ 		& 0.096 		& 0.075 		& 0.079 		& 0.068 		& 0.065 	\\ \hline
$E(FN>0)$ 		& 42.124 		& 33.732 		& 29.145		& 26.204 		& 23.865	\\ \hline
\end{tabular}
\caption{Type I and Type II error with $p=25$, $q=0.2$}
\end{table}

\begin{table}[h]
\centering
\begin{tabular}{| p{2cm} | p{0.8cm} | p{0.8cm} | p{0.8cm} | p{0.8cm} | p{0.8cm} | }

\hline
Bonferroni Step 	& 100 		& 200 		& 300 		& 400 		& 500 	\\ \hline
$E(P(FP>0))$ 		& 0.07 		& 0.062 		& 0.069 		& 0.062 		& 0.062	\\ \hline
$E(P(FN>0))$ 		& 1 			& 1 			& 1 			& 1 			& 1		\\ \hline
$E(FP>0)$ 		& 0.098 		& 0.08 		& 0.085 		& 0.073 		& 0.069 	\\ \hline
$E(FN>0)$ 		& 42.033 		& 33.658 		& 29.046		& 26.123 		& 23.803	\\ \hline
\end{tabular}
\caption{Type I and Type II error with $p=25$, $q=0.2$}
\end{table}

\begin{table}[h]
\centering
\begin{tabular}{| p{2cm} | p{0.8cm} | p{0.8cm} | p{0.8cm} | p{0.8cm} | p{0.8cm} | }

\hline
Sidak 			& 100 		& 200 		& 300 		& 400 		& 500 	\\ \hline
$E(P(FP>0))$ 		& 0.071 		& 0.065 		& 0.071 		& 0.063 		& 0.071	\\ \hline
$E(P(FN>0))$ 		& 1 			& 1 			& 1 			& 1 			& 1		\\ \hline
$E(FP>0)$ 		& 0.099 		& 0.084 		& 0.089 		& 0.075 		& 0.078 	\\ \hline
$E(FN>0)$ 		& 42.016 		& 33.59 		& 29.963		& 26.045 		& 23.734	\\ \hline
\end{tabular}
\caption{Type I and Type II error with $p=25$, $q=0.2$}
\end{table}

\begin{table}[h]
\centering
\begin{tabular}{| p{2cm} | p{0.8cm} | p{0.8cm} | p{0.8cm} | p{0.8cm} | p{0.8cm} | }

\hline
Sidak Step 		& 100 		& 200 		& 300 		& 400 		& 500 	\\ \hline
$E(P(FP>0))$ 		& 0.077 		& 0.07 		& 0.074 		& 0.065 		& 0.075	\\ \hline
$E(P(FN>0))$ 		& 1 			& 1 			& 1 			& 1 			& 1		\\ \hline
$E(FP>0)$ 		& 0.105 		& 0.089 		& 0.093 		& 0.08 		& 0.083 	\\ \hline
$E(FN>0)$ 		& 41.921 		& 33.507 		& 28.866		& 25.96 		& 23.666	\\ \hline
\end{tabular}
\caption{Type I and Type II error with $p=25$, $q=0.2$}
\end{table}

\begin{table}[h]
\centering
\begin{tabular}{| p{2cm} | p{1cm} | p{1cm} | p{1cm} | p{1cm} | p{1cm} | }

\hline
Sidak Step 		& 100 		& 200 		& 300 		& 400 		& 500 	\\ \hline
$E(P(FP>0))$ 		& 0.038 		& 0.048 		& 0.052 		& 0.041 		& 0.067	\\ \hline
$E(P(FN>0))$ 		& 1 			& 1 			& 1 			& 1 			& 1		\\ \hline
$E(FP>0)$ 		& 0.105 		& 0.089 		& 0.093 		& 0.08 		& 0.083 	\\ \hline
$E(FN>0)$ 		& 153.506		& 117.174		& 95.569		& 82.109 		& 73.635	\\ \hline
\end{tabular}
\caption{Type I and Type II error with $p=25$, $q=0.6$}
\end{table}

\begin{table}[h]
\centering
\begin{tabular}{| p{2cm} | p{1cm} | p{1cm} | p{1cm} | p{1cm} | p{1cm} | }

\hline
Sidak Step 		& 100 		& 200 		& 300 		& 400 		& 500 	\\ \hline
$E(P(FP>0))$ 		& 0.003 		& 0.008 		& 0.004 		& 0.005 		& 0.012	\\ \hline
$E(P(FN>0))$ 		& 1 			& 1 			& 1 			& 1 			& 1		\\ \hline
$E(FP>0)$ 		& 0.003 		& 0.008 		& 0.004 		& 0.005		& 0.012 	\\ \hline
$E(FN>0)$ 		& 263.477		& 219.876		& 191.627		& 171.954		& 157.58	\\ \hline
\end{tabular}
\caption{Type I and Type II error with $p=25$, $q=0.95$}
\end{table}

Tables 2,3,4 and 5 show 4 different adjustment procedures for matrix with $p=25$ and $q=0.2$. It means that there are 300 possible connections, and about 60 true edges. As a result, there can be not more than 60 Type II errors in total. Tables 6 and 7 show Sidak step-down adjustment procedure for different densities of matrices. Experiments showed that:
\begin{itemize}[noitemsep]
\item Achieved practical level of FWER depends on density of a matrix for any of the procedures
\item The number of Type II error is particularly high in case of greater dimensionality even for significant number of observations (for 7x7 case this number is usually ~0 for 500 observations)
\item The number of Type II errors is a bit smaller for step procedures, the number of Type I errors is also slightly bigger for step procedures, however the difference is small in comparison to the total number of errors, or wrongly defined connections.
\end{itemize}

Other experiments compared ROC AUC for the four procedures with adjustments and simultaneous multiple testing procedure. ROC AUC measure allows us to compare these procedures without looking at their significance level. The experiments were conducted for the matrix with $p=7$, which has 9 non-zero elements from the interval $[0.2, 0.55]$.
\begin{figure}
\includegraphics[scale = 0.3]{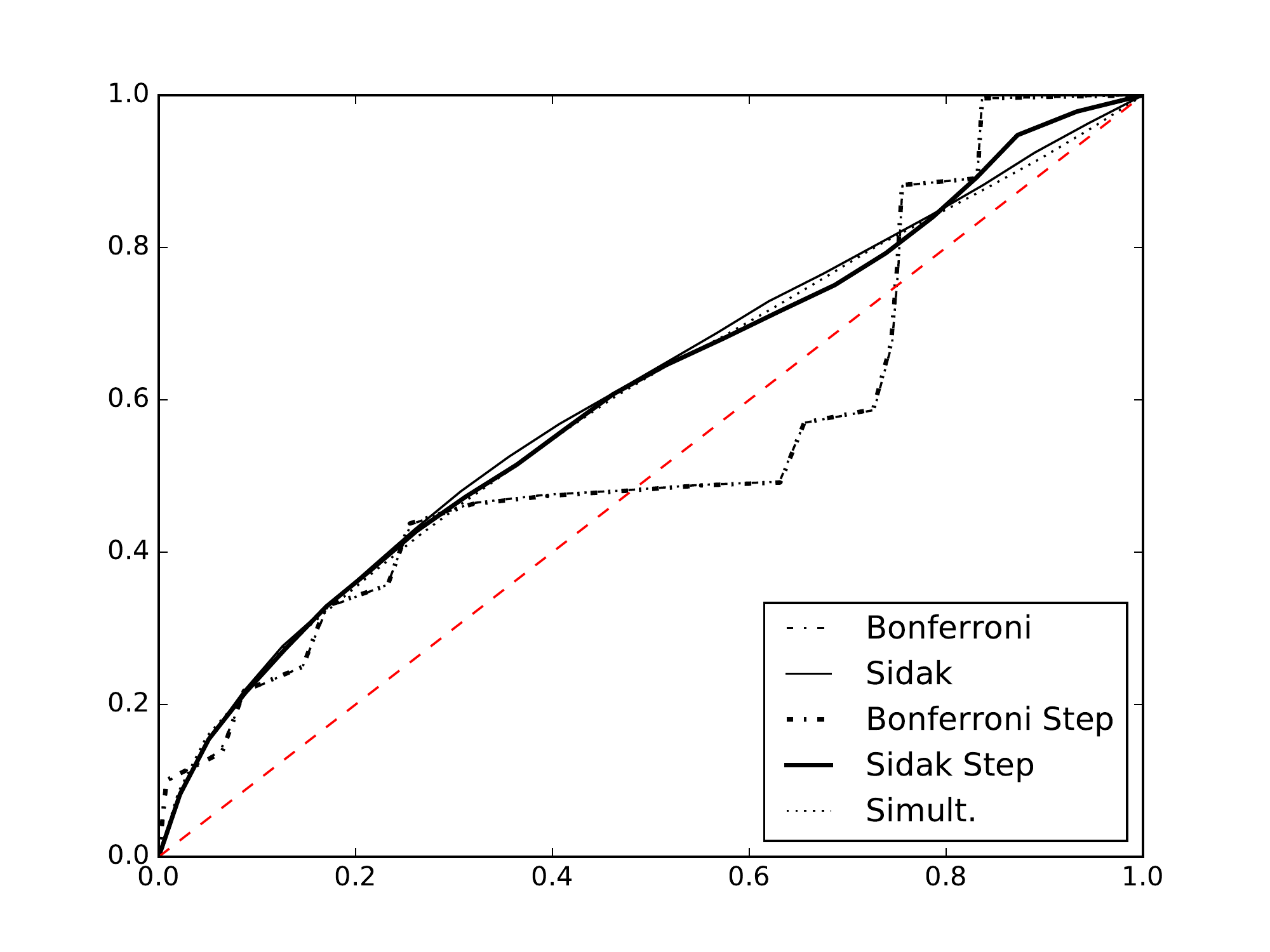}
\includegraphics[scale = 0.3]{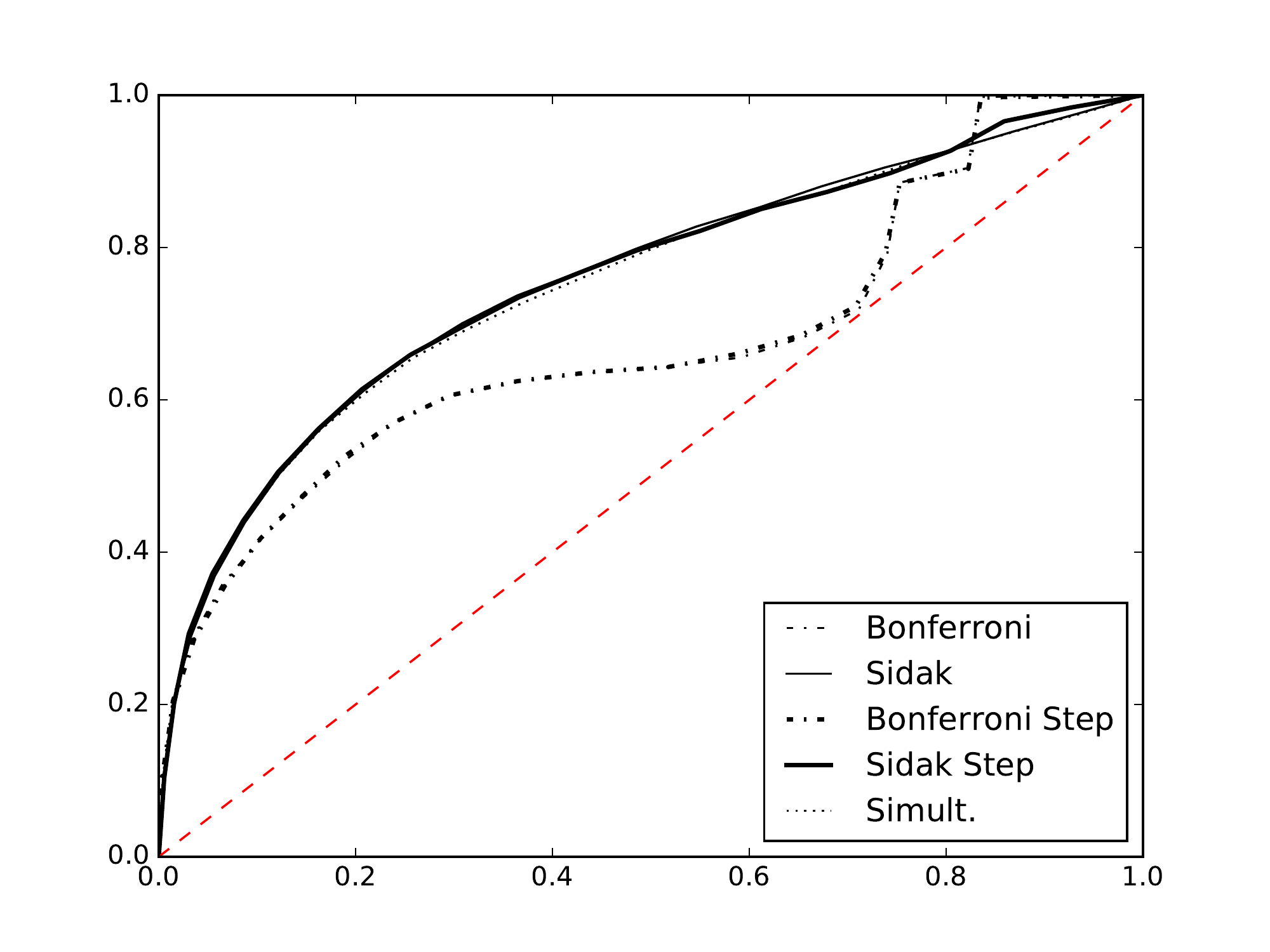}
\caption{ROC for different procedures, $n = 10, 20$}
\end{figure}
\begin{figure}
\includegraphics[scale = 0.3]{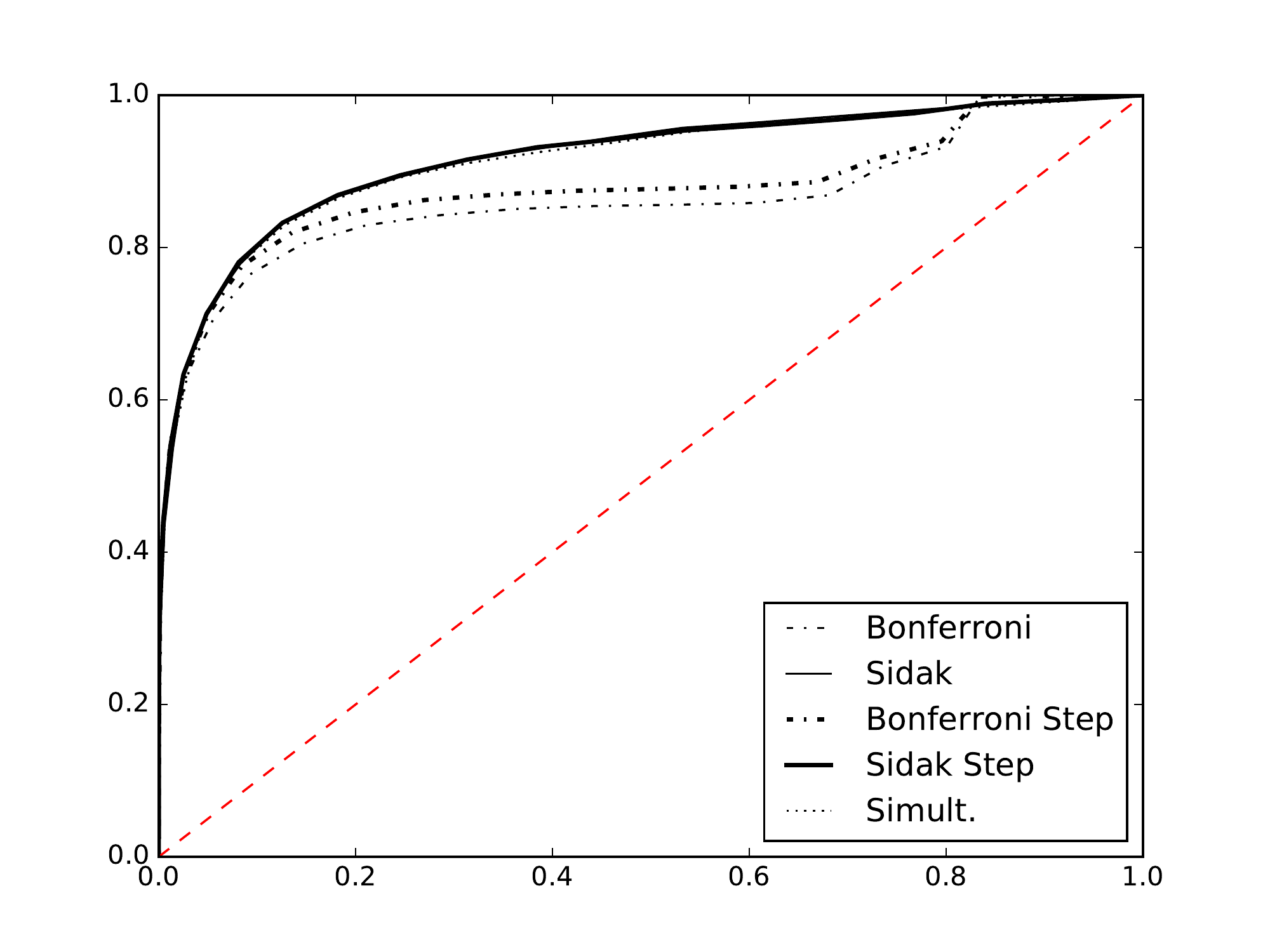}
\includegraphics[scale = 0.3]{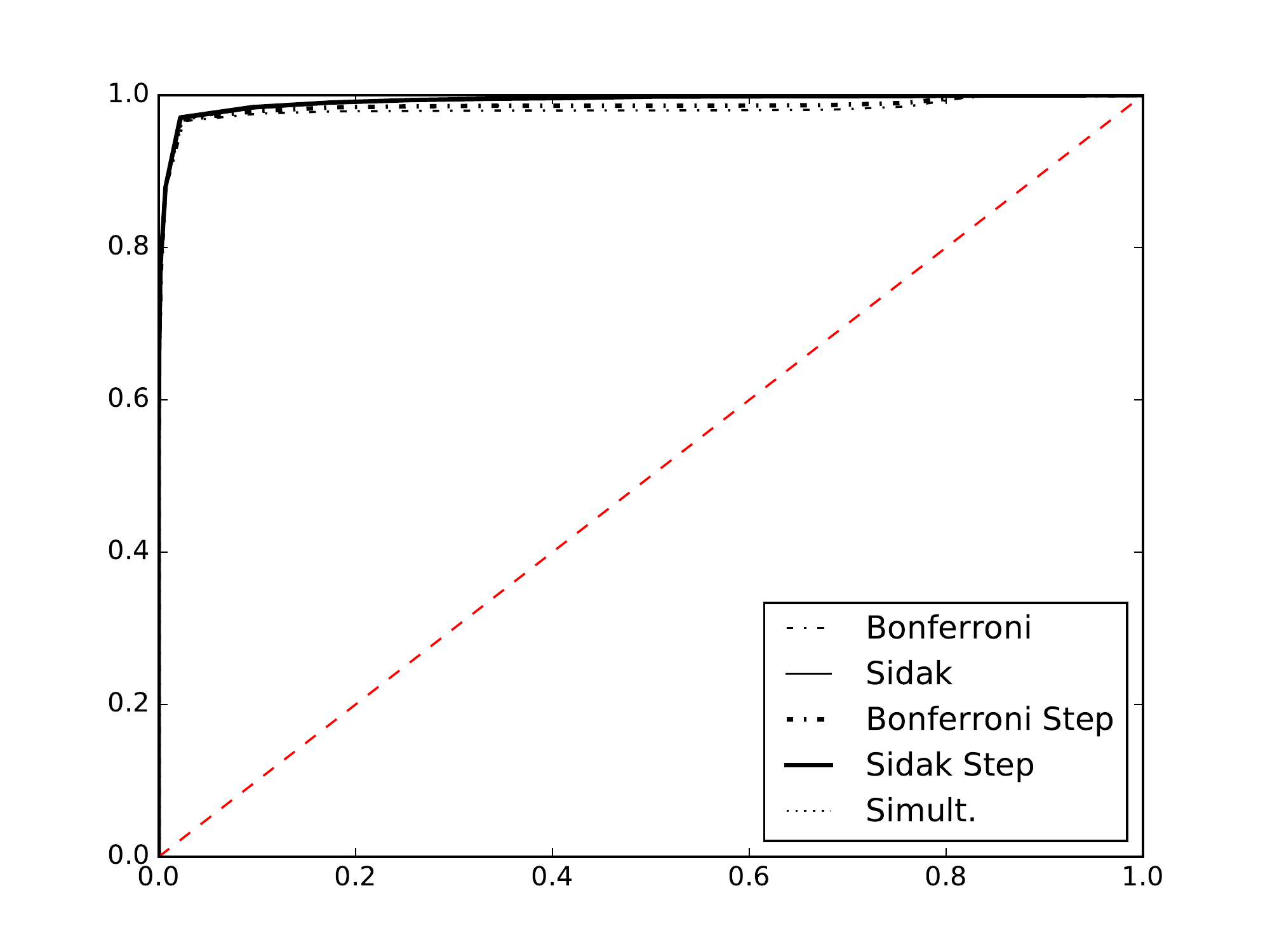}
\caption{ROC for different procedures, $n = 50, 150$}
\end{figure}

Increasing number of observations led to increasing ROC AUC, which can be clearly seen from figures 2 and 3. The best ROC AUC is achieved for Sidak adjustment and simultaneous multiple testing procedure without adjustments. Differences between those three procedures are insignificant. 

Finally, we compare risk functions for different procedures. On the picture, horizontal axis is different values of $\alpha$ and vertical axis is values of risk function.

\begin{figure}
\includegraphics[scale = 0.3]{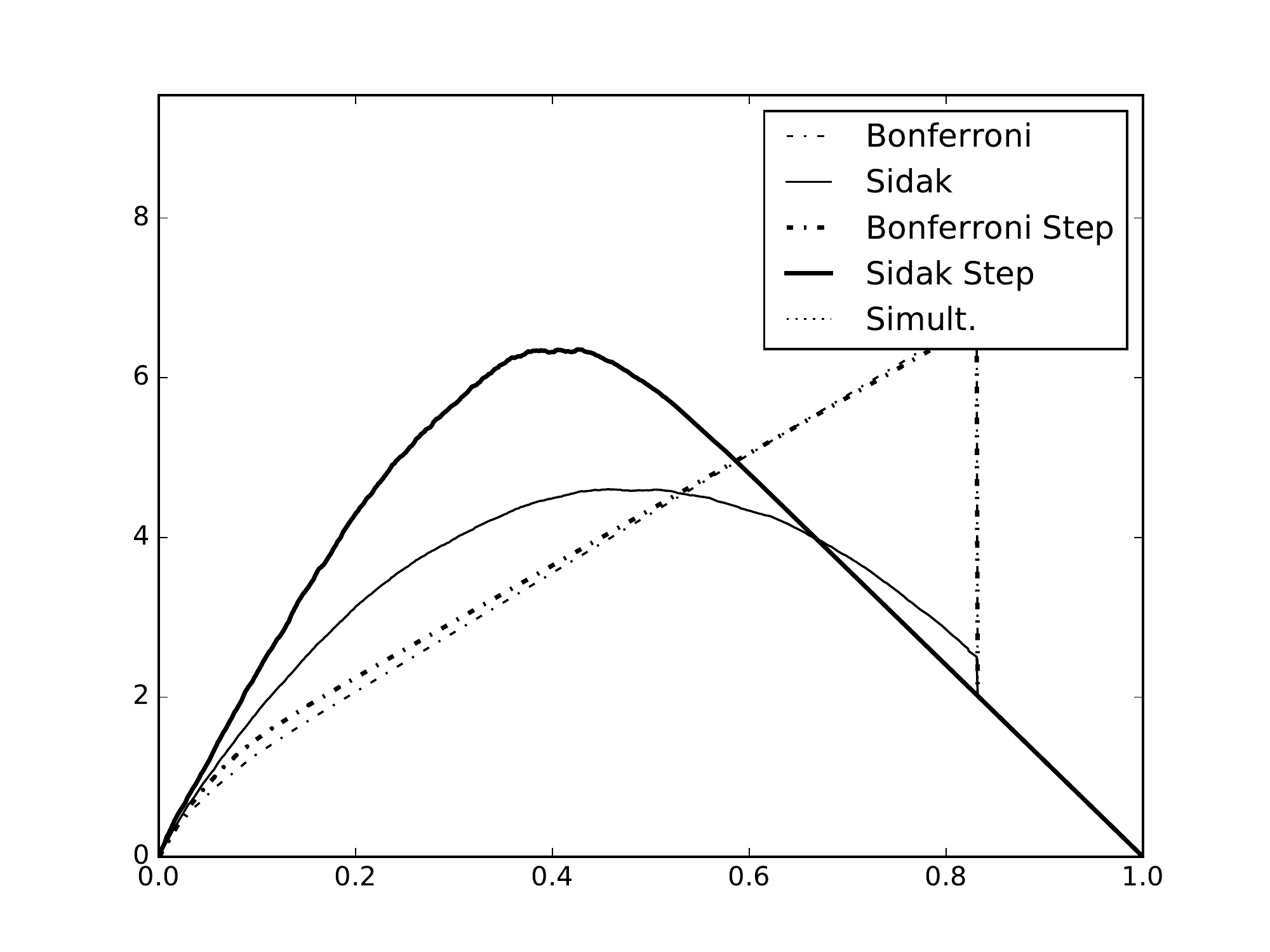}
\includegraphics[scale = 0.3]{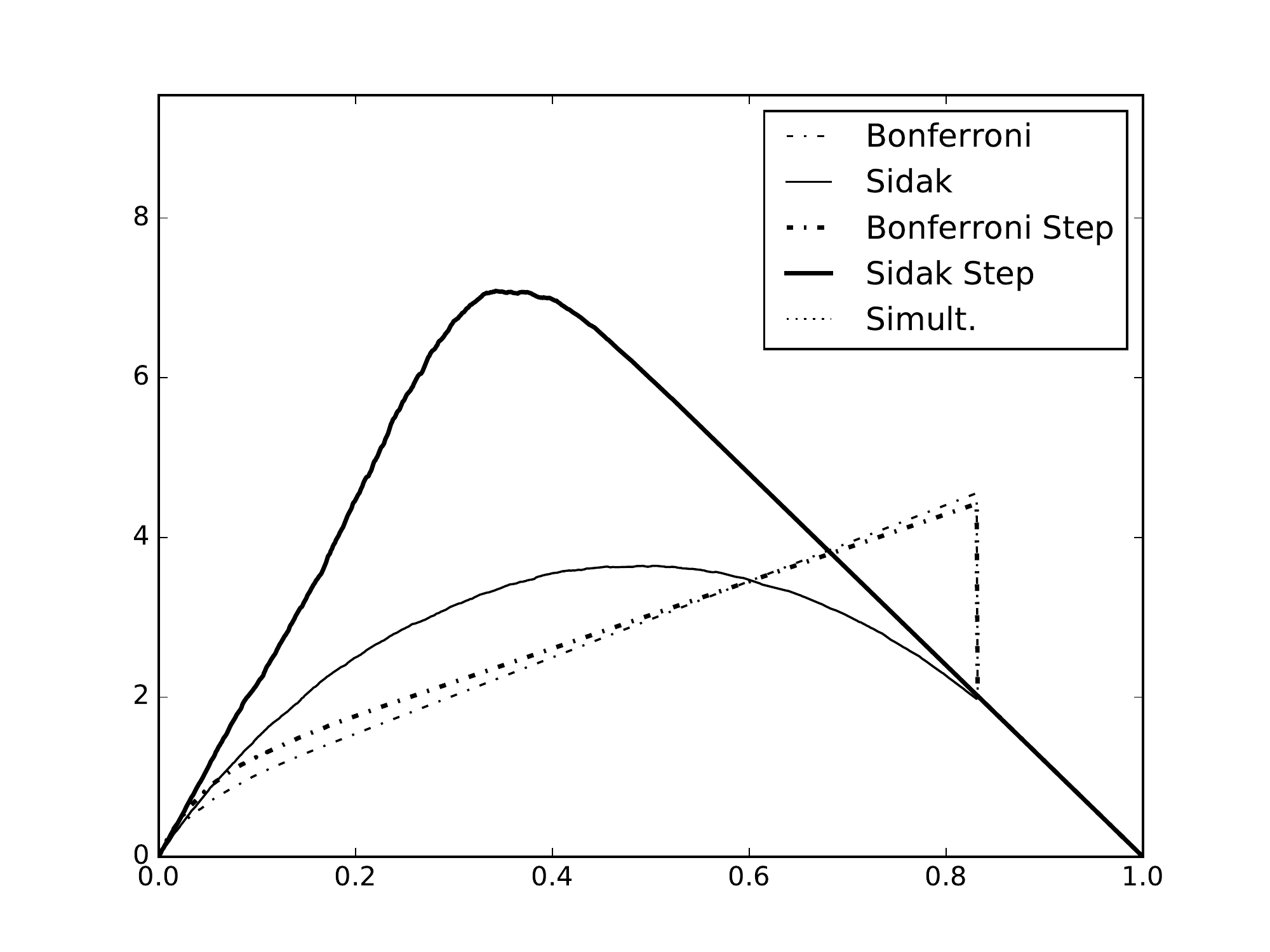}
\caption{Risk function for different procedures, $n = 10, 20$}
\end{figure}
\begin{figure}
\includegraphics[scale = 0.3]{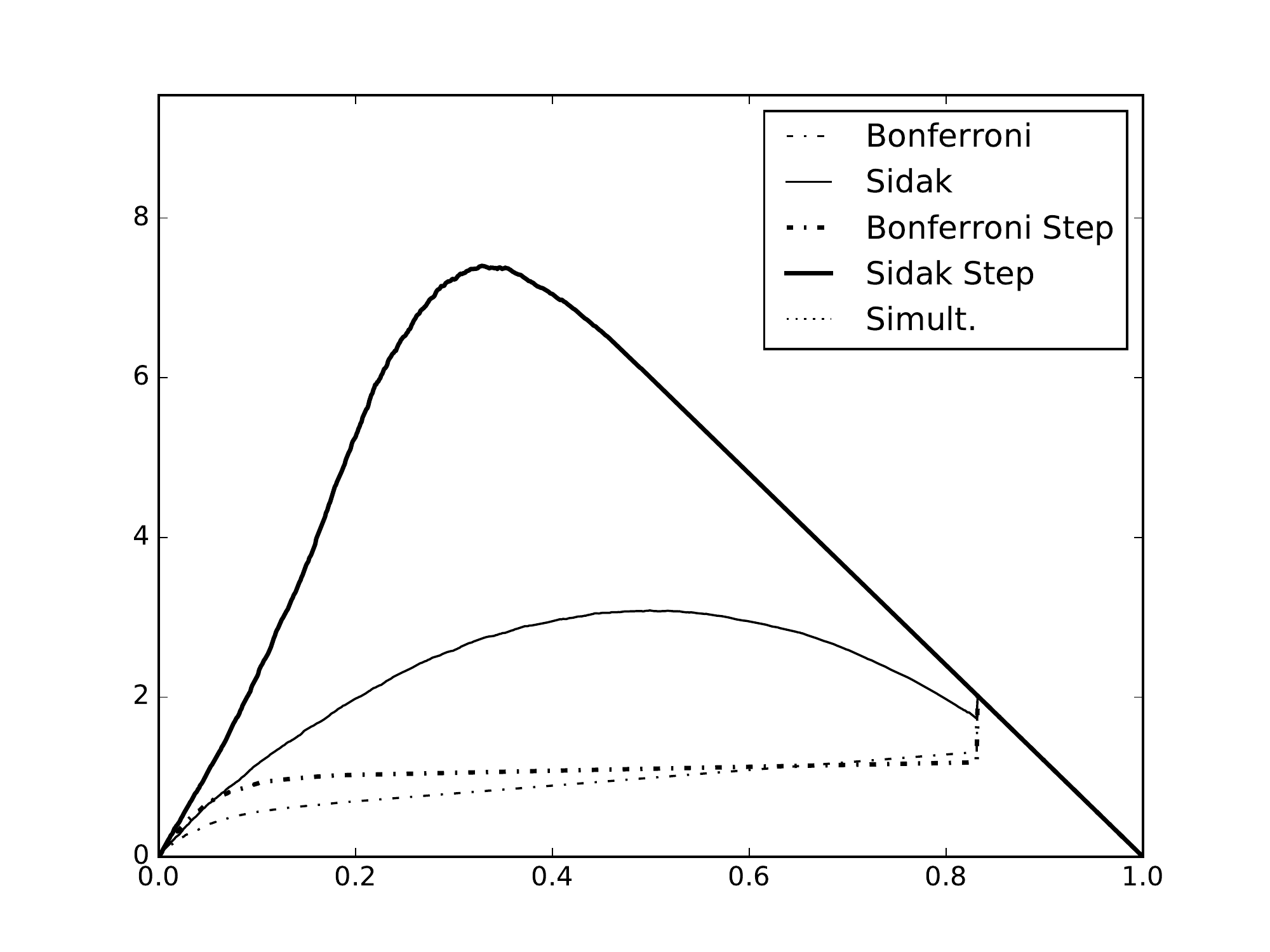}
\includegraphics[scale = 0.3]{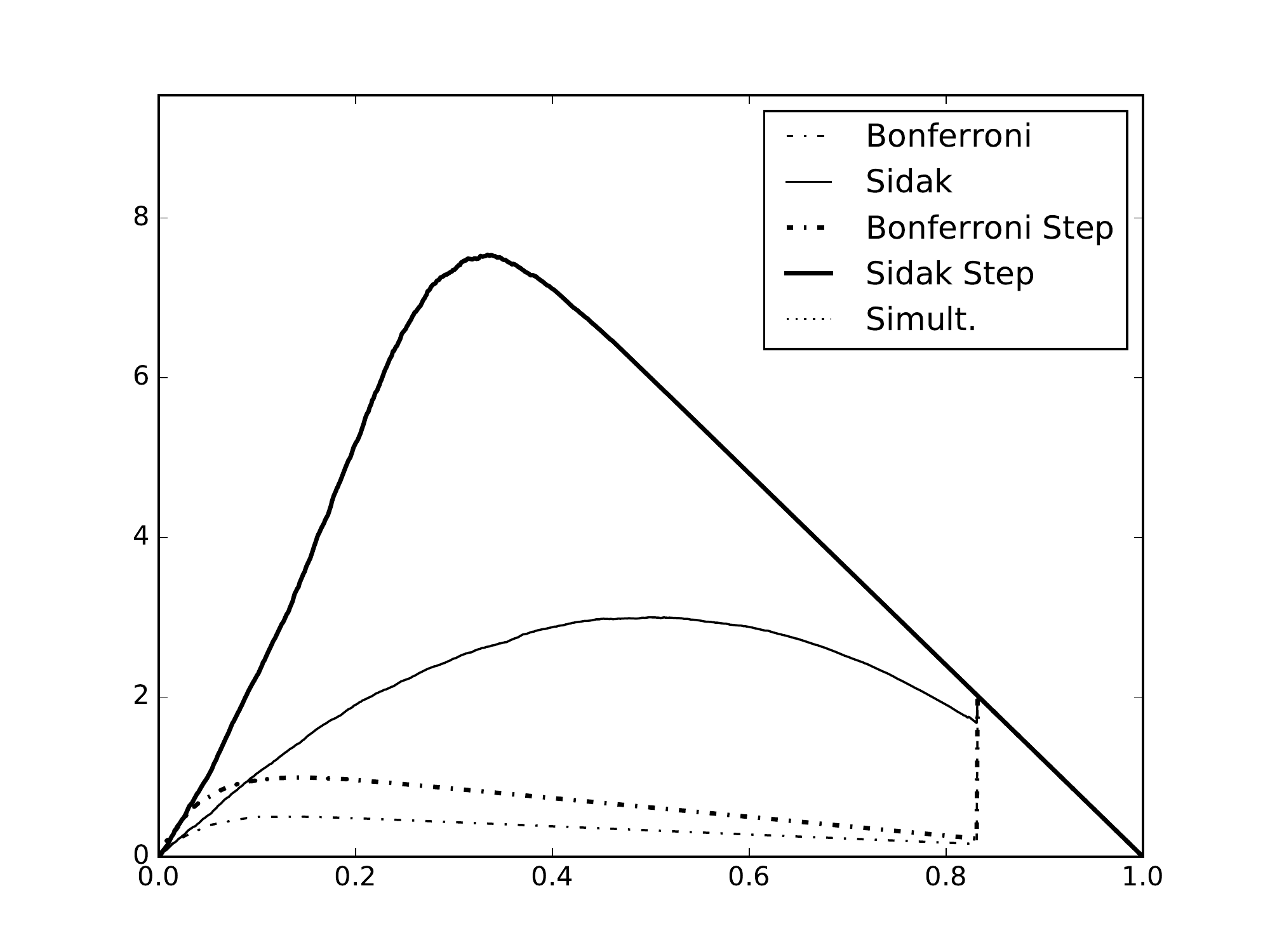}
\caption{Risk function for different procedures, $n = 50, 140$}
\end{figure}

According to figures 4 and 5, risk function for simultaneous multiple testing coincides with risk function for Sidak step-down adjustment procedure. Additionally, the value of risk function for these two procedures increases with the number of observations, whereas Bonferroni, Bonferroni step-down and Sidak adjustment lower the value of risk function with rising number of observations.

\section{Conclusion}

In the article we analyzed procedures, described in Drton \& Perlman from some new points of view. Despite the control of FWER for these procedures, they perform poorly from the point of view of Type II errors. Sidak, Sidak step-down adjustment procedures and simultaneous multiple testing procedure show similar ROC curves with almost equal AUC score, which improves with growing number of observations. As a result, Sidak, Sidak step-down adjustment procedures procedures and simultaneous multiple testing procedure may be considered best amongst analyzed, however some of their properties are still not satisfactory. The directions for the future work include analysis of goodness-of-fit procedures and research of properties of observed procedures for other elliptical distributions.

\section{Acknowledgments}

This work was conducted at the Laboratory of Algorithms and Technologies for Network Analysis.

\pagebreak

\end{document}